\documentclass[pra,twocolumn,showpacs,preprintnumbers,amsmath,amssymb,superscriptaddress,unsortedaddress]{revtex4}
\usepackage{graphicx}% Include figure files
\usepackage{dcolumn}% Align table columns on decimal point
\usepackage{bm}% bold math

\begin{document}

\preprint{}

\title{Increasing singlet fraction with entanglement swapping}% Force line breaks with \\

\author{Joanna Mod{\l}awska}

\affiliation{Faculty of Physics, Adam Mickiewicz University, ul. Umultowska 85, 61-614 Pozna\'{n}, Poland}%Lines break automatically or can be forced with \\

\author{Andrzej Grudka}

\affiliation{Faculty of Physics, Adam Mickiewicz University, ul. Umultowska 85, 61-614 Pozna\'{n}, Poland}

\affiliation{Institute of Theoretical Physics and Astrophysics,
University of Gda\'{n}sk, ul.  Wita Stwosza 57, 80-952 Gda\'{n}sk, Poland}

\affiliation{National Quantum Information Centre of Gda\'{n}sk, ul. W{\l}adys{\l}awa Andersa 27, 81-824 Sopot, Poland}

\date{\today}% It is always \today, today,

\begin{abstract}
We consider entanglement swapping for certain mixed states. We assume that the initial states have the same singlet fraction and show that the final state can have singlet fraction greater than the initial states. We also consider two quantum teleportations and show that entanglement swapping can increase teleportation fidelity. Finally, we show how this effect can be demonstrated with linear optics.
\end{abstract}

\maketitle

Maximally entangled states are crucial for quantum information processing. If two parties share a pair of qubits in the maximally entangled state then they can perform quantum teleportation. However, usually the parties share nonmaximally entangled state $\varrho$. One can define the singlet fraction $F$ of the state $\varrho$ as the maximal overlap of the state $\varrho$ with the maximally entangled state, i.e.,
\begin{equation}
F=\text{max} \langle \Psi | \varrho | \Psi \rangle,
\end{equation}
where the maximum is taken over all maximally entangled states $| \Psi \rangle$. If one performs quantum teleportation with the state $\varrho$ preprocessed by local unitary operations \cite{Horodecki5} then the optimal teleportation fidelity is
\begin{equation}
f=\frac{2F+1}{3}.
\end{equation}
If $F > \frac{1}{2}$ then the parties can perform quantum teleportation with the average fidelity of the teleported qubit exceeding classical limit $\frac{2}{3}$. However, it is well known that there are two-qubit entangled states which have $F < \frac{1}{2}$. The Horodecki family has proved that one can increase the singlet fraction of any two-qubit entangled state above $\frac{1}{2}$ by non-trace preserving local operations and classical communication (LOCC) \cite{Horodecki2}. Verstraete and Verschelde have proved that one can do it even with trace preserving LOCC \cite{Verstraete1}. Moreover, they have found how to obtain optimal teleportation with any two-qubit state, i.e., how to find an LOCC protocol which gives the highest average fidelity of the teleported qubit. Let us now consider a string of nodes connected by nonmaximally entangled pairs of qubits. The first set of entangled pairs is distributed between nodes $A$ and $B$, the second one is distributed between $B$ and $C$ and so on. In order to perform quantum teleportation from the first to the last node one first distills entanglement between $A$ and $B$, then between  $B$ and $C$ and so on \cite{Bennett3}. Next, one performs entanglement swapping \cite{Zukowski1, Zukowski2} at each node which creates entanglement between the first and the last node. Notice that usually many entangled pairs are needed between two nodes in order to distill entanglement. Finally, one performs quantum teleportation between the first and the last node. This strategy is crucial ingredient of quantum repeaters \cite{Briegel}. However, for pure nonmaximally entangled states there exists another strategy. Instead of distilling maximally entangled state between each pair of neighboring nodes and then performing entanglement swapping one first performs entanglement swapping and then distills entanglement between the first and the last node \cite{Bose1, Bose2, Hardy3, Acin, Perseguers1}. If each pair of neighboring nodes is connected by a single nonmaximally entangled pure state this strategy gives higher probability of obtaining maximally entangled state between the first and the last node.

In this paper we consider entanglement swapping for certain mixed states. We assume that we have two pairs of nonmaximally entangled states with the same singlet fraction. We perform entanglement swapping and show that for certain initial states the singlet fraction of the final state is greater than the singlet fraction of the initial states. In particular, we show that the singlet fraction of the initial states can be smaller than $\frac{1}{2}$ and the singlet fraction of the final state can be greater than $\frac{1}{2}$. Thus the initial pairs of qubits are not useful for quantum teleportation (if one does not first increase their singlet fractions) and the final pair of qubits is useful for quantum teleportation.
This effect does not occur for all mixed states. For example it was shown that if one performs entanglement swapping with mixtures of two Bell states or Werner states, then the final state has always the singlet fraction less than the initial states \cite{Sen1, Sen2}. We will also consider maximization of the fidelity of quantum teleportation. We show that the fidelity of two teleportations is below $\frac{2}{3}$ if we maximize the fidelity of each teleportation independently. However, if we first perform entanglement swapping and then teleport a qubit, then the fidelity of quantum teleportation is above $\frac{2}{3}$.

Before we present mixed states for which one can increase the singlet fraction with entanglement swapping we discuss for which states one cannot do that. In Ref. \cite{Kent1,Linden1} it was proved that the parties who use LOCC cannot increase the singlet fraction of a \emph{single} copy of entangled Bell-diagonal state (e.g. Werner states or mixtures of two Bell states). We assume that two parties  (Alice and Bob) share such a state. Let one party (Bob) prepare in his laboratory additional \emph{arbitrary} entangled state. We stress that this additional state is held by Bob and it is not shared by Alice and Bob. Next, Bob performs Bell measurement on a particle from the state which he shares with Alice and on the first particle from his additional entangled state. It is clear that the state of Alice's particle and the second particle from Bob's additional entangled state cannot have the singlet fraction greater than the singlet fraction of the original state because Bob performed only local operations. We conclude that one cannot increase the singlet fraction of entangled Bell-diagonal states with entanglement swapping. 

Now, we present mixed states for which one can increase the singlet fraction with entanglement swapping. We assume that Alice and Bob as well as Bob and Charlie share a pair of qubits in mixed entangled state. Alice and Bob share a state
\begin{eqnarray}
& \varrho_{AB}=p (\sqrt{a}|0\rangle |1\rangle - \sqrt{1-a}|1\rangle |0\rangle)(\langle 0| \langle 1|\sqrt{a} -\nonumber \\
& - \sqrt{1-a}\langle 1| \langle 0|)+ (1-p) |0\rangle |0 \rangle \langle 0| \langle 0|
\end{eqnarray}
and Bob and Charlie share a state
\begin{eqnarray}
& \varrho_{BC}=p (\sqrt{a}|1\rangle |0\rangle - \sqrt{1-a}|0\rangle |1\rangle)(\langle 1| \langle 0|\sqrt{a} -\nonumber \\
& - \sqrt{1-a}\langle 0| \langle 1|)+ (1-p) |0\rangle |0 \rangle \langle 0| \langle 0|.
\end{eqnarray}
Both states are entangled for $p \neq 0$ and $0<a<1$ and have the same singlet fraction
\begin{equation}
F=\text{max}\{p(\sqrt{a}+\sqrt{1-a})^2/2,(1-p)/2 \}.
\end{equation}
These states are mixtures of nonmaximally entangled state and an orthogonal product state and can be obtained by sending one qubit from an entangled pair of qubits through amplitude damping channel. Let Bob measures his qubits from two mixed states, which he shares with Alice and Charlie, in the Bell basis $\{|\Psi^{\pm}\rangle, |\Phi^{\pm}\rangle\}$, where
\begin{equation}
|\Psi^{\pm}\rangle=\frac{1}{\sqrt{2}}(|0\rangle |1\rangle \pm |1\rangle |0\rangle)
\end{equation}
and
\begin{equation}
|\Phi^{\pm}\rangle=\frac{1}{\sqrt{2}}(|0\rangle |0\rangle \pm |1\rangle |1\rangle).
\end{equation}
If Bob obtains $|\Psi^{+}\rangle$ or $|\Psi^{-}\rangle$ as the result of his measurement, then Alice's and Charlie's qubits are found in the state (after local unitary operation)
\begin{eqnarray}
& \varrho_{AC}=\frac{1}{N}(2p^2 a(1-a) |\Psi^{-}\rangle \langle \Psi^{-}|+\nonumber \\
& +2p(1-p)a|0\rangle |0\rangle \langle 0| \langle 0|)
\end{eqnarray}
and $N=2p^2 a (1-a)+2p(1-p)a$ is the probability of obtaining $|\Psi^{+}\rangle$ or $|\Psi^{-}\rangle$ as the result of Bob's measurement. The singlet fraction of this state is
\begin{equation}
F=\text{max}\{ \frac{2p^2 a(1-a) }{N}, \frac{p(1-p)a}{N} \}.
\end{equation}
\begin{figure}
\includegraphics{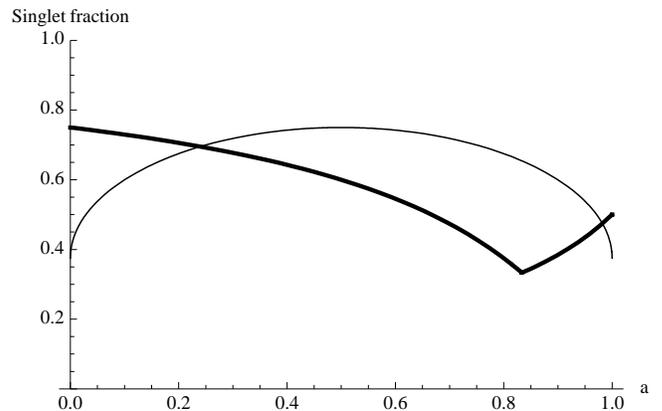}
\caption{\label{fig:1} Singlet fraction versus parmeter $a$ for $p=0.75$ for the initial states (thin line) and for the final state (thick line)  if Bob obtains $|\Psi^{\pm}\rangle$  as the result of his measurement.}
\end{figure}

In Fig. 1 we present how the singlet fraction of the initial states and the singlet fraction of the final state depend on the parameter $a$ for $p=0.75$ when Bob obtains $|\Psi^{+}\rangle$ or $|\Psi^{-}\rangle$ as the result of his measurement.  For $a<0.0285955$ the initial states have the singlet fraction less than $\frac{1}{2}$ and for $a<0.666667$ the final state has the singlet fraction greater than $\frac{1}{2}$.  Hence, for $a<0.666667$, if Alice teleports a qubit to Charlie with the final state then the fidelity of the teleported qubit will exceed the classical limit. For  $a>0.666667$, the final state is still entangled because it is a mixture of maximally entangled state and an orthogonal product state but one has first to increase its singlet fraction in order to perform quantum teleportation and exceed the classical limit. 

If Bob obtains $|\Phi^{+}\rangle$ or $|\Phi^{-}\rangle$ as the result of his measurement then Alice's and Charlie's qubits are found in the state (after local unitary operation)
\begin{eqnarray}
& \varrho_{AC}=\frac{2}{1-N}(p^2(a |0\rangle|1\rangle - (1-a)|1\rangle|0\rangle) \nonumber\\
& (a \langle0|\langle1| - (1-a)\langle1|\langle0|)+(1-p)^2|0\rangle |1\rangle \langle 0| \langle 1|\nonumber \\
& +p(1-p)(1-a)(|0\rangle |0\rangle \langle 0| \langle 0|+|1\rangle |1\rangle \langle 1| \langle 1|)
\end{eqnarray}
and $1-N=1-2p^2 a (1-a)-2p(1-p)a$ is the probability of obtaining $|\Phi^{+}\rangle$ or $|\Phi^{-}\rangle$ as the result of Bob's measurement. The singlet fraction of this state is:
\begin{eqnarray}
& F=\text{max}\{ \frac{1 - 2 p + 2 p^2}{2(1-N)}, \nonumber\\
& \frac{(-1 + a) (-1 + p) p}{1-N}\}.
\end{eqnarray}
\begin{figure}
\includegraphics{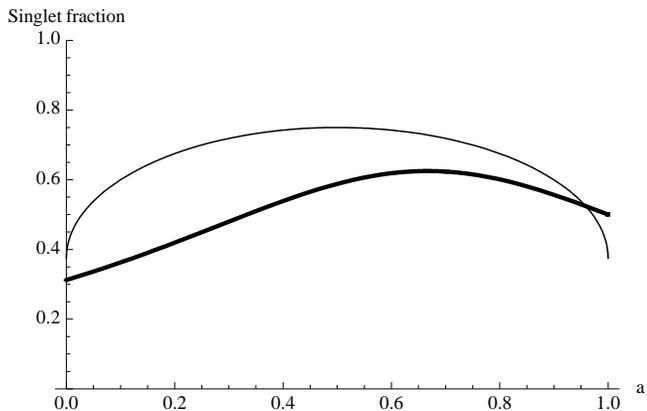}
\caption{\label{fig:2} Singlet fraction versus parmeter $a$ for $p=0.75$ for the initial states (thin line) and for the final state (thick line) if Bob obtains $|\Phi^{\pm}\rangle$  as the result of his measurement.}
\end{figure}

In Fig. 2 we present how the singlet fraction of the initial states and the singlet fraction of the final state depend on the parameter $a$ for $p=0.75$ when Bob obtains $|\Phi^{+}\rangle$ and $|\Phi^{-}\rangle$ as the result of his measurement. For $a<0.333333$ the final state has the singlet fraction less than $\frac{1}{2}$ and moreover, it is separable.

So far we have shown how one can increase the singlet fraction above $\frac{1}{2}$ with entanglement swapping  probabilistically, i.e., by non-trace preserving operations, if the initial states have the singlet fraction below $\frac{1}{2}$. Now we show how one can increase the singlet fraction above $\frac{1}{2}$ with entanglement swapping deterministically, i.e., by trace preserving operations. We use similar idea as in Ref. \cite{Verstraete1}. Let Bob measures his qubits in Bell basis. We suppose that if Bob obtains $|\Psi^{\pm}\rangle$ as the result of his measurement (which happens with probability $N$), then Alice and Charlie have a state with the singlet fraction greater than $\frac{1}{2}$. Let maximum in Eq. (1) for Alice's and Charlie's state be obtained for maximally entangled state $|\Psi^{-}\rangle$. If Bob obtains  $|\Phi^{\pm}\rangle$  as the result of his measurement (which happens with probability $1-N$) and Alice and Bob have a state with the singlet fraction less than $\frac{1}{2}$, then they prepare a product state $|01\rangle$. Hence, on average the singlet fraction is greater than $\frac{1}{2}$.

Let us now assume that Alice wants to send quantum information to Bob.  We consider two statrategies. The first strategy is as follows. Alice and Bob increase deterministically the singlet fraction of their entangled state and Alice teleports a qubit to Bob. Next, Bob and Charlie increase deterministically the singlet fraction of their entangled state and Bob teleports a qubit (which he received from Alice) to Charlie. We assume that the parties try to optimize each teleportation independently. It was shown \cite{Verstraete1} that in the optimal trace preserving LOCC protocol maximizing the singlet fraction for state $\rho$ one party applies a local filter and classically communicates the result to the other party. If the filtering succeeds the parties do nothing. Otherwise they prepare pure product state.  One can find the optimal filter $A^{*}$ and the fidelity $F^{*}$ by solving the following semidefinite program: maximize
\begin{equation}
F=\frac{1}{2}-\text{Tr}(X \rho^{\Gamma})
\end{equation}
under the constraints:
\begin{equation}
0 \leq X \leq I_{4},
\end{equation}
\begin{equation}
-\frac{I_{4}}{2} \leq X^{\Gamma} \leq \frac{I_{4}}{2}.
\end{equation}
and $X$ is of rank $1$. The filter $A$ is related to $X$ by the following equation
\begin{equation}
X=I_{2}\otimes A|\Phi^{+}\rangle \langle \Phi^{+}|I_{2} \otimes A^{\dagger}.
\end{equation}
In order to solve the semidefininte program we observe that the state $\rho_{AB}^{\Gamma}$ (and the state $\rho_{BC}^{\Gamma}$) has the following symetries \cite{Verstraete1}: it is invariant under transposition and local operations $\sigma_{z} \otimes \sigma_{z}$ and $\text{diag}[1,i] \otimes \text{diag}[i, 1]$. The operator $X$ has to have the same symmetries and hence it is of the form
\begin{equation}
X=\left( \begin{array} {cccc}  x_1 & 0 & 0 & x_2\\
0 & x_3 & 0 & 0\\
0 & 0 & x_4 & 0\\
x_5 & 0 & 0 & x_6
\end{array} \right)
\end{equation}
with $x_1, x_2, x_3, x_4, x_5, x_6$ real. Moreover, since $X$ is of rank one $x_3$ and $x_4$ have to be equal to $0$. Now maximization can be easily done. We obtain the following expression for the maximal singlet fraction and local filter:
\begin{equation}
F^{*}=\frac{1}{2}+\frac{a(1-a)p^2}{2(1-p)},
\end{equation}
\begin{equation}
A^{*}=\left( \begin{array} {cc} \frac{\sqrt{a} \sqrt{1-a} p}{1-p} & 0\\
0 & 1
\end{array} \right),
\end{equation}
if $\frac{\sqrt{a} \sqrt{1-a} p}{1-p} < 1$;
\begin{equation}
F^{*}=\frac{p}{2}+\sqrt{a(1-a)}p,
\end{equation}
\begin{equation}
A^{*}=\left( \begin{array} {cc} 1 & 0\\
0 & 1
\end{array} \right),
\end{equation}
if $\frac{\sqrt{a} \sqrt{1-a} p}{1-p} \geq 1$.
The final state of Alice and Bob (as well as Bob and Charlie), after they applied trace preserving operation to the initial state, is:
\begin{eqnarray}
& \rho_{AB}^{*}=I_{2}\otimes A^*\rho_{AB}I_{2}\otimes A^{*\dagger}+ \nonumber\\
& +(1-\text{Tr}(I_{2}\otimes A^*\rho_{AB}I_{2}\otimes A^{*\dagger}))|0\rangle |1\rangle \langle 0| \langle 1|,
\end{eqnarray}

\begin{eqnarray}
& \rho_{BC}^{*}=A^*\otimes I_{2}\rho_{BC} A^{*\dagger} \otimes I_{2}+ \nonumber\\
& +(1-\text{Tr}(A^*\otimes I_{2}\rho_{BC} A^{*\dagger}\otimes I_{2}))|1\rangle |0\rangle \langle 1| \langle 0|.
\end{eqnarray}

Now, Alice teleports a qubit to Bob with entangled state of Eq. (21) and then Bob teleports a qubit to Charlie with entangled state of Eq. (22). 

The second strategy is as follows. Bob performs entanglement swapping. If after Bob's measurement Alice and Charlie have a state with the singlet fraction greater than $\frac{1}{2}$, then they transform it with local unitary operations in such a way  that the maximal overlap of the state with maximally entangled state is obtained for a state $|\Psi^{-}\rangle$.  If after Bob's measurement they have a state with the singlet fraction less than $\frac{1}{2}$, then they prepare a product state $|01\rangle$. Note that this is not optimal strategy because if the singlet fraction is below $\frac{1}{2}$ and Alice's and Charlie's state is entangled, then they can increase the singlet fraction above $\frac{1}{2}$. Hence, for the initial states of Eqs. (3) and (4) and $p=0.75$ the final state of Alice and Charlie and its singlet fraction is:
\begin{widetext}
1) for $p=0.75$ and $a \leq 0.333333$

\begin{equation}
\rho_{AC}=\left( \begin{array} {cccc}  -2 a (-1 + p) p & 0 & 0 & 0\\
0 & 1 + a (-2 + p) p + a^2 p^2 & a (-1 + a) p^2 & 0\\
0 & a (-1 + a) p^2 & -a (-1 + a) p^2 & 0\\
0 & 0 & 0 & 0
\end{array} \right),
\end{equation}

\begin{equation}
F=\frac{1}{2} - a^2 p^2 + a p (-1 + 2 p),
\end{equation}

2) for $p=0.75$ and $0.333333<a<0.666667$

\begin{equation}
\rho_{AC}=\left( \begin{array} {cccc}  -(1 + a) (-1 + p) p & 0 & 0 & 0\\
0 & 1 - 2 p + (1 + a) p^2 & 2 a (-1 + a) p^2 & 0\\
0 & 2 a (-1 + a) p^2 & -(-1 + a) p^2 & 0\\
0 & 0 & 0 & (-1 + a) (-1 + p) p
\end{array} \right),
\end{equation}

\begin{equation}
F=\frac{1}{2} - p + (1 + 2 a - 2 a^2) p^2,
\end{equation}

3) for $p=0.75$ and $a \geq 0.666667$

\begin{equation}
\rho_{AC}=\left( \begin{array} {cccc}  (-1 + a) (-1 + p) p & 0 & 0 & 0\\
0 & 1 + 2 (-1 + a) p - (-1 + a^2) p^2 & a (-1 + a) p^2 & 0\\
0 & a (-1 + a) p^2 & (-1 + a)^2 p^2 & 0\\
0 & 0 & 0 & (-1 + a) (-1 + p) p
\end{array} \right),
\end{equation}

\begin{equation}
F=\frac{1}{2} + (-1 + a) p - (-1 + a^2) p^2.
\end{equation}

\end{widetext}

\begin{figure}
\includegraphics {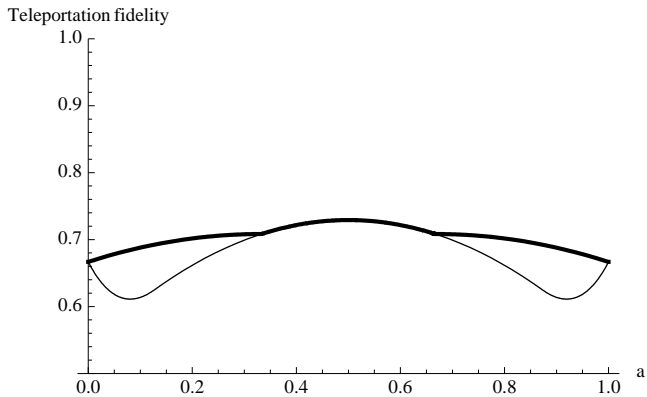}
\caption{\label{fig:3} Teleportation fidelity versus parmeter $a$ for $p=0.75$ for two subsequent teleportations with independently filtered states (thin line) and for one teleportation with state resulting from entanglement swapping (thick line).}
\end{figure}

In Fig. 3 we present how teleportation fidelity for two strategies depends on the parameter $a$ for $p=0.75$. One can see that if Bob first performs entanglement swapping, then Alice can teleport a qubit to Charlie with teleportation fidelity greater than $\frac{2}{3}$ for $0<a<1$.  On the other hand if the parties increase the singlet fractions of the two states independently, then for $a<0.211325$ and $a>0.788675$ teleportation fidelity does not exceed classical limit. For $0.333333<a<0.666667$ the two strategies give the same teleportation fidelity. This is connected to the fact that for $0.333333<a<0.666667$ the final state resulting from entanglement swapping has the singlet fraction greater than $\frac{1}{2}$ for each result of Bob's measurements and the two strategies are equivalent. If $a<0.333333$ or $a>0.666667$ and Bob performs entanglement swapping, then the final state has the singlet fraction greater  than $\frac{1}{2}$ for two results of Bob's measurement and less than $\frac{1}{2}$ for the other two results of Bob's measurements. In such a case Alice and Charlie replace it with a product state which has the singlet fraction equal to $\frac{1}{2}$.

The effect which we described can be demonstrated experimentally with linear optics, single photon sources and photodetectors discriminating the number of photons. First one prepares two entangled states
\begin{eqnarray}
& |\Psi^{(AB)}\rangle=\sqrt{1-p(1-a)}|0\rangle_a |1\rangle_{b_1} -\nonumber \\
&  - \sqrt{p(1-a})|1\rangle_a |0\rangle_{b_1}
\end{eqnarray}
and
\begin{eqnarray}
& |\Psi^{(BC)}\rangle=\sqrt{1-p(1-a)}|1\rangle_{b_2} |0\rangle_c -\nonumber \\
& - \sqrt{p(1-a)}|0\rangle_{b_2} |1\rangle_c
\end{eqnarray}
where $|i\rangle_{a}$ stands for $i$ photons in mode $a$ and similarly for other modes. Beam splitters with transmission coefficients $T=\frac{pa}{1-p(1-a)}$ are placed in modes $b_1$ and $b_2$ (see Fig. 3). They realize amplitude damping channels and produce the states of Eqs. 2 and 3. The photons from modes $b_{3}$ and $b_{4}$ are not detected. The modes $b_1$ and $b_2$ are incoming modes of beam splitter with transmission coefficient $T=0.5$. After this beam splitter there are photodetectors which discriminate the number of photons.  If only one of two photodetectors detects one photon  then the result of the measurement is $|\Psi^+\rangle_{b_1b_2}$ or $|\Psi^-\rangle_{b_1b_2}$ and for appropriately chosen parameters $a$ and $p$ one increases the singlet fraction.

\begin{figure}
\includegraphics [width=9truecm]{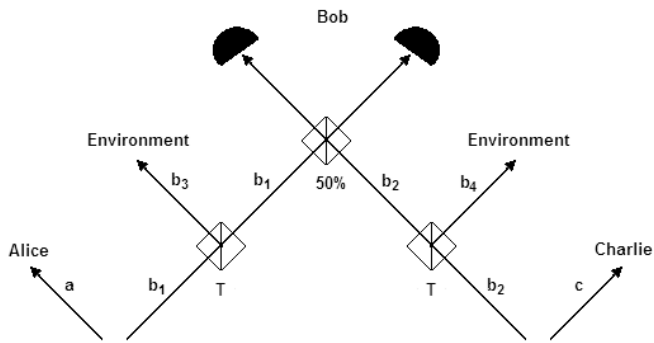}
\caption{\label{fig:4} Experimental setup}
\end{figure}

In conclusion we have shown that entanglement swapping can increase the singlet fraction. For some states the singlet fraction of the initilal states can be less than $\frac{1}{2}$ and the singlet fraction of the final state can be greater than $\frac{1}{2}$. We have also considered the usefulness of entanglement swapping in quantum teleportation. Finally, we have shown how this effect can be demonstrated experimentally. Our result may have applications in quantum networks.

\begin{acknowledgments}
We thank Micha{\l} Horodecki, Pawe{\l} Horodecki, Ryszard Horodecki and Marek \.{Z}ukowski for discussions. One of the authors (A.G.) was partially supported by Ministry of Science and Higher
Education Grant No. N N206 2701 33 and by the European Commission through the Integrated Project FET/QIPC SCALA.
\end{acknowledgments}

\end{document}